\documentclass[notitlepage,superscriptaddress,showpacs,nobalancelastpage,aps,prl,twocolumn]{revtex4-1}
\usepackage[english]{babel}
\RequirePackage[T1]{fontenc}

\usepackage{amsfonts}
\usepackage{subfigure}
\usepackage{amsmath}
\usepackage{amssymb}
\usepackage{graphicx,epstopdf}
\usepackage{array}
\usepackage[hidelinks]{hyperref}
\usepackage{color}
\usepackage{my_symbols}
\usepackage{CJK}
\usepackage{bm}
\usepackage{tikz}

\begin{document}
\begin{CJK*}{UTF8}{gbsn} 
\title{Manipulating spin wave polarization in synthetic antiferromagnet}
\date{\today}

\author{Jin Lan (兰金)}
\affiliation{Department of Physics and State Key Laboratory of Surface Physics, Fudan University, Shanghai 200433, China}
\author{Weichao Yu (余伟超)}
\affiliation{Department of Physics and State Key Laboratory of Surface Physics, Fudan University, Shanghai 200433, China}
\author{Jiang Xiao (萧江)}
\email[Corresponding author:~]{xiaojiang@fudan.edu.cn}
\affiliation{Department of Physics and State Key Laboratory of Surface Physics, Fudan University, Shanghai 200433, China}
\affiliation{Collaborative Innovation Center of Advanced Microstructures, Nanjing 210093, China}
\affiliation{Institute for Nanoelectronics Devices and Quantum Computing, Fudan University, Shanghai 200433, China}

\begin{abstract}
Polarization is a key ingredient of all waves, including the electromagnetic wave, the acoustic wave, as well as the spin wave.
Due to the fixed ferromagnetic order, the spin wave in ferromagnet is limited to the right circular polarization.
The spin wave in antiferromagnet, however, is endowed with the full polarization degree of freedom because of the two identical magnetic sublattices.
In the synthetic antiferromagnet, the two magnetic sublattices are spatially separated into two sublayers.
The circular polarization of spin wave is partially locked to the magnetic sublattice of the antiferromagnet, thus to the sublayer in synthetic antiferromagnet.
Based on this unique polarization-sublayer locking mechanism, we show that
both the circular spin wave polarizer and retarder (wave-plate) can be straightforwardly realized using synthetic antiferromagnets by restructuring the sublayers, e.g. by removing or capping a portion of a sublayer.
Manipulating spin wave polarization by geometrical engineering provides a simple yet powerful paradigm in harnessing the spin wave polarization for spin information processing.
\end{abstract}

\maketitle
\end{CJK*}

\emph{Introduction.}
{As the collective excitation of the magnetic order, spin wave is a promising candidate for next-generation energy-saving information carrier because it can propagate devoid of Joule heating \cite{kruglyak_magnonics_2010,lenk_building_2011,chumak_magnon_2015}.}
Similar as other types of wave, the spin wave also possesses the polarization degree of freedom, which represents the precessing direction of the magnetic order \cite{tveten_antiferromagnetic_2014,cheng_antiferromagnetic_2016,lan_antiferromagnetic_2017}.
In ferromagnet, such polarization degree of freedom is frozen to be the right-circular polarization
with respect to the given magnetization direction.
However in antiferromagnet, spin wave can have both left and right circular polarizations, thus the full polarization degree of freedom, because of the existence of two identical magnetic lattices with opposite magnetization \cite{kittel_theory_1951,keffer_theory_1952,duine_synthetic_2018}.

It is much more convenient to use wave polarization, instead of wave amplitude or phase, to encode or process information \cite{Goldstein_2003_polarized,sophia_phonon_2014,sklan_splash_2015}.
In ideal antiferromagnet with two identical magnetic sublattices, spin waves with different polarizations are degenerate in energy.
To manipulate spin wave polarization, one must lift this degeneracy, for example, by applying external magnetic field, or by introducing Dzyaloshinskii-Moriya interaction (DMI) \cite{dzyaloshinsky_thermodynamic_1958,moriya_anisotropic_1960}.
Utilizing DMI, the spin wave version of Faraday effect \cite{cheng_antiferromagnetic_2016} and spin Nernst effect \cite{cheng_spin_2016,zyuzin_magnon_2016}, as well as the spin wave polarizing and retarding effect
\cite{lan_antiferromagnetic_2017} have been proposed.

Synthetic antiferromagnet (SyAF), consisting of two magnetic sublayers coupled antiferromagnetically \cite{lavrijsen_magnetic_2013,yang_domain_2015}, mimics the real antiferromagnet in many aspects, including the spin wave behaviors and its polarization properties.
In real antiferromagnet, the two magnetic sublattices are intimately interweaved thus are difficult to manipulate each of them separately or differentially.
In contrast, the two magnetic sublattices in an SyAF are spatially separated into two sublayers, which makes it possible to control the magnetic sublattices independently or differentially.
The recently discovered $2$D van der Waals magnetic materials
\cite{huang_layer-dependent_2017,gong_discovery_2017,xing_electric_2017,wang_raman_2016,lee_ising-type_2016} can also serve as a special type of SyAF with two layers of 2D magnetic materials coupled antiferromagnetically.

In this Rapid Communication, we propose to
manipulate the spin wave polarizations
in SyAF by engineering the magnetic sublayer.
 Specifically, removing a portion of one sublayer in a SyAF realizes a spin wave polarizer in the basis of circular polarization; while capping a portion of the SyAF  with another sublayer  realizes a spin wave retarder (wave-plate), also in the basis of circular polarization. Harnessing the spin wave polarization by sublayer engineering in SyAF offers a new designing principle in controlling the spin waves in the circular polarization basis.

\begin{figure*}[tp]
\includegraphics[width=0.98\textwidth, trim=0 380 0 0,clip ]{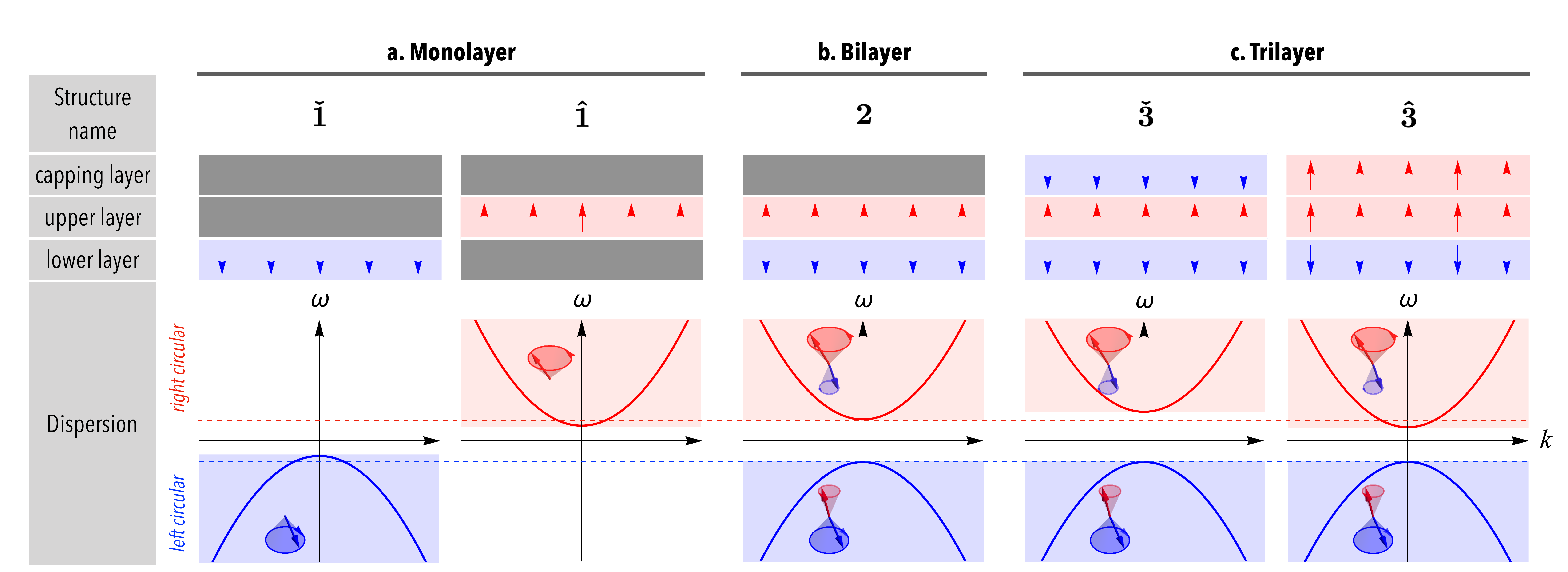}
\caption{ {\bf Spin waves in synthetic antiferromagnet and its modifed structures}.
The magnetic structure and the corresponding spin wave dispersions in (a) monolayer, (b) bilayer, (c) trilayer synthetic antiferromagnet.
Structures:
The static magnetizations in $\pm\hbz$ direction are denoted by red (blue) arrows. The monolayer/trilayer structure is constructed from a bilayer synthetic antiferromagnet by removing/capping one sublayer.
The monolayer with magnetization pointing downward/upward is denoted as $\check{1}/\hat{1}$ structure, and the trilayer with magnetization in the capping layer pointing downward/upward is denoted as $\check{3}/\hat{3}$ structure.
Dispersions:
The dispersions for the right/left circular modes are plotted as the positive/negative branch. In bilayer and trilayer, the right/left circular modes are preferentially locallized in the upper/lower sublayer with larger precession amplitude. The dashed lines denote the spin wave gap for the bilayer.
 }
\label{fig:pols}
\end{figure*}

\emph{Model.} Consider a bilayer SyAF with two ferromagnetic sublayers coupled antiferromagnetically as depicted in \Figure{fig:pols}(b), in which blue (red) arrows denote the magnetization directions of the lower and upper sublayer.
The magnetization dynamics in the SyAF is governed by two coupled Landau-Lifshitz-Gilbert (LLG) equations \cite{yang_domain_2015,lan_antiferromagnetic_2017},
\begin{equation}
\dbm_j(\br,t) = -\gamma\mb_j(\br,t)\times \bH_j^\ssf{eff}+ \alpha \mb_j(\br,t) \times \dbm_j(\br,t),
\label{eqn:LLG}
\end{equation}
where $j=1,2$ denotes two sublayers, $\gamma$ is the gyromagnetic ratio, and $\alpha$ is the Gilbert damping constant.
Here $\gamma \bH_j^\ssf{eff}(\br,t) = K m_j^z\hbz+ A \nabla^2 \mb_j- J \mb_{\bar{j}}+\gamma H_j\hbz$ (with $\bar{1} = 2$ and $\bar{2} = 1$) is the effective magnetic field acting locally on sublattice $\mb_j(\br)$, where $K$ is the easy-axis anisotropy (along $\hbz$), $A$ is the intra-layer ferrromagnetic exchange coupling constant within each sublayer, $J$ is the inter-layer antiferromagnetic exchange coupling constant between two sublayers, and $H_j$ is the external magnetic field (in $\hbz$) acting on sublayer-$j$.

\emph{Spin wave polarization and magnetic sublayer.}
At equilibrium, the sublayer magnetization $\mb_{1/2}^0$ is in $\mp \hbz$ direction respectively.
Upon this collinear magnetic configuration, we separate the static and dynamical components of the sublayer magnetization as $\mb_j(\br,t) = \mb_j^0+ \delta \mb_j(\br,t)$, where $\delta \mb_j= m_j^x\hbx+m_j^y \hby$ is the dynamical transverse component of the excited spin wave upon the static magnetization $\mb_j^0$.
By eliminating the uniform magnetization background from the LLG equation (\ref{eqn:LLG}), the spin wave dynamics reduces to the following linearized equations:
\begin{equation}
\label{eqn:EOM_afm}
(-)^j i \partial_t m_j^-
= \smlb{-A \nabla^2+K +J+\gamma H_j} m_j^- + J m_{\bar{j}}^-,
\end{equation}
where $\partial_t \equiv \partial/\partial t$, $m_j^- \equiv m_j^x- i m_j^y$, and $j = 1, 2$.
Without external field ($H_j = 0$),
\Eq{eqn:EOM_afm} hosts two degenerated spin wave modes of opposite circular polarizations with dispersion $\omega_\ssf{R/L} = \pm \sqrt{(A k^2+K+J)^2-J^2}$, which locates in the negative/positive branch of the dispersion respectively, as shown in \Figure{fig:pols}(b).

{A key feature about the circularly polarized spin wave in antiferromagnet is that the magnetization in both sublattices precesses, but with different precession amplitudes \cite{kittel_theory_1951}.
Specifically in SyAF, with two sublattices spatially separated into two sublayers, the precession amplitudes in two sublayers would be different.
As described in \Eq{eqn:EOM_afm}, the right (left) circular spin wave mode is preferentially localized on the upper (lower) sublayer, i.e. the circular polarization of spin wave is closely connected to magnetic sublayer in SyAF.
}


\emph{Manipulating magnetic sublayer.}
Because of the polarization-sublayer locking feature in SyAF, it is straightforward to manipulate polarization via sublayer engineering.
The simplest sublayer operation is to remove or cap one sublayer of an SyAF, as shown in \Figure{fig:pols}(a)(c).
As a result, a bilayer SyAF becomes the a monolayer or trilayer magnetic structure, respectively.

By removing the upper/lower sublayer in \Figure{fig:pols}(b), the bilayer SyAF becomes the trivial monolayer ferromagnet with magnetization pointing downward/upward, denoted as the $\check{1}/\hat{1}$ structure as shown in \Figure{fig:pols}(a).
The spin wave dynamics in the $\check{1}$ and $\hat{1}$ structure are governed by \Eq{eqn:EOM_afm} with $J = \gamma H_j =0$, where the interlayer coupling is turned off in the bilayer SyAF.
As a result, the $\check{1}$/$\hat{1}$ structure only accommodates the left/right circularly polarized mode, which is associated with the lower/upper sublayer in the bilayer SyAF.
Correspondingly, the spin wave dispersion in $\check{1}$/$\hat{1}$ structure is single-branched: $\omega_\ssf{R/L}= \mp (K+A k^2)$, being the left and right circularly polarized respectively as depicted in \Figure{fig:pols}(a).

Alternatively, by capping a third ferromagnetic layer, the bilayer SyAF becomes a trilayer structure as shown in \Figure{fig:pols}(c). The capping layer is antiferromagnetically coupled to the upper sublayer of the SyAF with the same interlayer exchange coupling $J$.  Depending on the magnetization direction (down or up) of the capping layer, and the resulting trilayer is denoted as $\check{3}$ or $\hat{3}$ structure.
Assuming the magnetization in the capping layer is pinned along $\hbz$: $\mb_3=\mp \hbz$
\footnote{The spin wave behaviors with pinned/free $\mb_3$ are similar (see Supplementary Materials), therefore in the following we focus on the pinned $\mb_3$ case. },
the magnetization dynamics in the trilayer SyAF is described by \Eq{eqn:EOM_afm} with $\gamma H_1=0$ and $\gamma H_2 = -J m_3^z=\pm J$ for the $\check{3}/\hat{3}$ structure.
The magnetization of the upper sublayer in the trilayer is subjected to the additional effective magnetic field ($\gamma H_2 \neq 0$) from the neighbouring capping layer, while that of the lower sublayer ($\gamma H_1 = 0$) is unaffected.
Consequently in the $\check{3}/\hat{3}$ structure, the dispersion of the right circular spin wave mode, which is preferentially localized in the upper layer, is substantially modified to $\omega_\ssf{R} = \pm J/2 + \sqrt{ (Ak^2+K+J\pm J/2)^2-J^2 }$; while the dispersion of the left circular mode, which is preferentially residing in the lower sublayer, is only slightly changed as $\omega_L= \pm J/2 -\sqrt{ (Ak^2+K+J\pm J/2)^2-J^2 }$, as shown in \Figure{fig:pols}(c).

\begin{figure}[tb]
\includegraphics[width=0.48\textwidth,trim=40 5 15 5,clip]{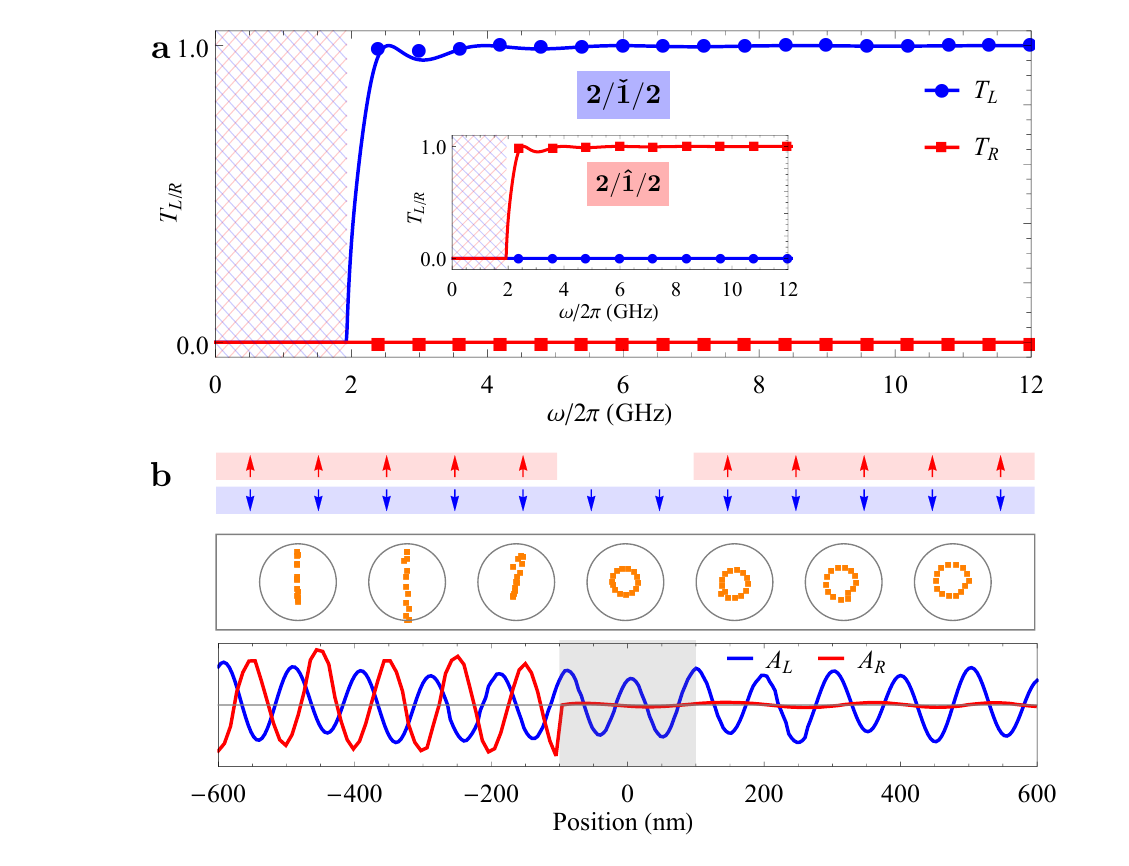}
\caption{{\bf Circular spin wave polarizer in the $2/1/2$ structure.}
{\bf a}. The polarization-dependent spin wave transmission probabilities $T_\ssf{L/R}$ across the $2/\check{1}/2$ structure (main panel) and $2/\hat{1}/2$ (inset).
The solid lines are calculated from Green function, and the dots are extracted from micromagnetic simulations.
The hatched area indicates that the frequency is below the gap of the left (blue) and right (red) circularly polarized modes.
{\bf b}. Micromagnetic simulation of a circular spin wave polarizer based on the $2/\check{1}/2$ structure.
Linearly polarized spin wave is injected from the left side with frequency $\omega/2\pi=6.5~\mathrm{GHz}$.
Top: the $2/\check{1}/2$ structure with a $200$ nm long central $\check{1}$ region;
Middle: the Lissajous-like plot of staggered order $n_x$ and $n_y$;
Bottom: the amplitude for left/right circular component $A_\ssf{L/R}$ as a function of position.
}
\label{fig:212}
\end{figure}

\begin{figure*}[tb]
\includegraphics[width=0.95\textwidth,trim= 0  5 15 25,clip]{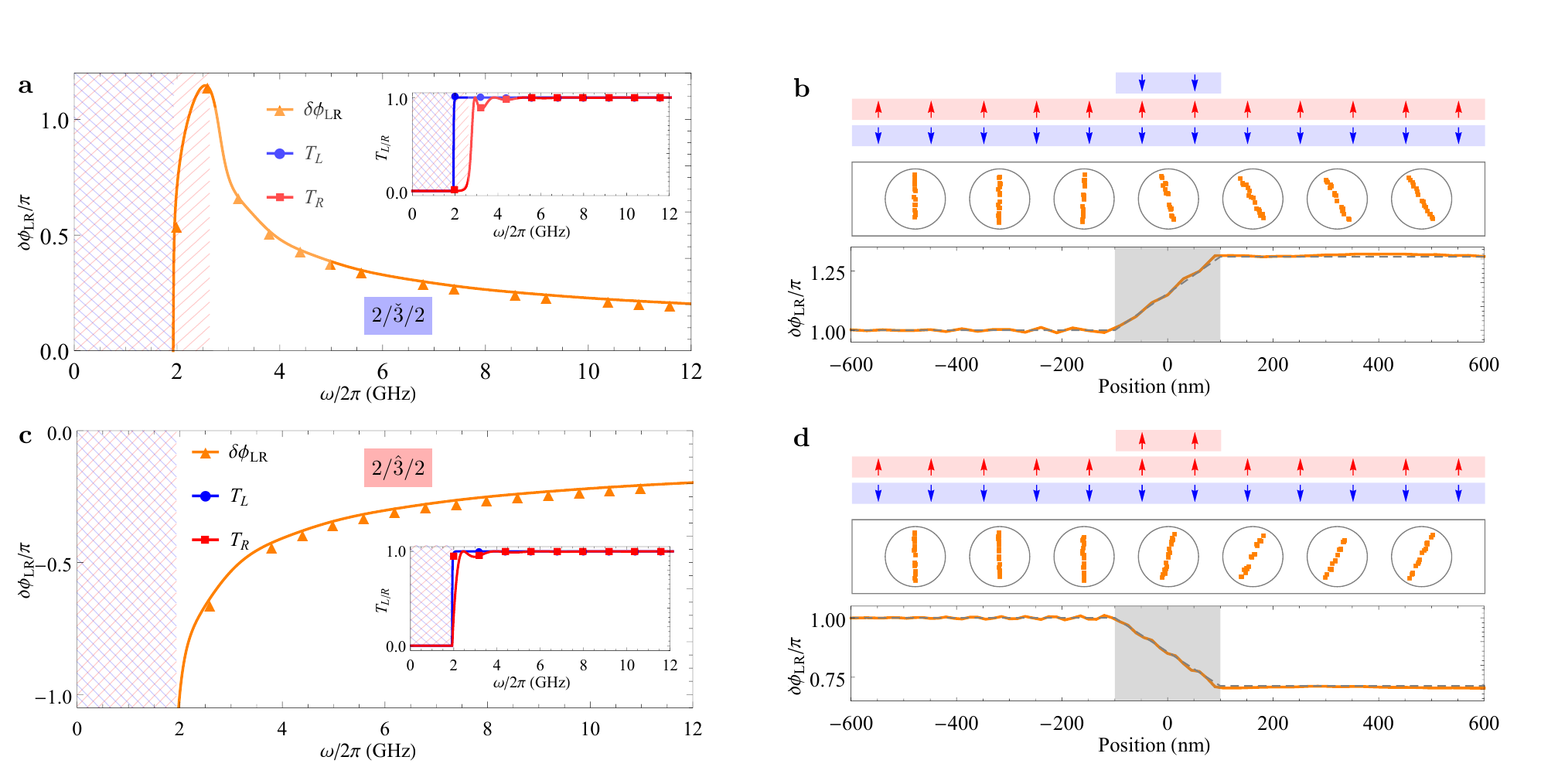}
\caption{ {\bf Circular spin wave retarder in the $2/3/2$ structure.}
{\bf a.} The relative phase $\delta \phi_\ssf{LR}$ (main panel) and the transmission probabilities $T_\ssf{L/R}$ (inset) as function of the spin wave frequency for the $2/\check{3}/2$ structure.
The solid lines are calculated from Green function, and the dots are extracted from micromagnetic simulations.
{\bf b.} Micromagnetic simulations of a spin wave circular retarder based on the the $2/\check{3}/2$ structure.
Linear-$y$ spin wave is injected from the left side with frequency $\omega/2\pi= 6.5~\mathrm{GHz}$.
Top: the $2/\check{3}/2$ structure with a $200$ nm long central $\check{3}$ region;
Middle: the Lissajous-like plot of the staggered order $n_x$ and $n_y$;
Bottom: the relative phase $\delta \phi_\ssf{LR}$ as a function of position extracted from micromagnetic simulations.
{\bf c., d.} Same as {\bf a., b.}, but for the $2/\hat{3}/2$ structure.
}
\label{fig:retarder}
\end{figure*}

As presented above, the degeneracy between left and right circular spin wave in an SyAF can be lifted by either sublayer-removal or sublayer-capping. The former operation eliminates one of the circular polarizations completely, and the latter modifies the dispersions asymmetrically. In terms of conducting polarized spin waves, the uncompensated monolayer, fully compensated bilayer, and partially compensated trilayer structure mimic the ``half-metal'', ``normal metal'', and ``ferromagnetic metal'' for spin-polarized conduction electrons, respectively \cite{zutic_spintronics_2004}.
It is well known in conventional spintronics that rich spin manipulation can be realized by using magnetic heterostructures using various combinations of normal metal, ferromagnetic metal, and half-metal. In a similar fashion, to manipulate polarized spin waves, one can also construct the effective magnetic heterostructures by various combiantions of the monolayer, bilayer, and trilayer SyAFs.
The simplest example of all are the $2/1/2$ and $2/3/2$ structures as shown in \Figure{fig:212} and \Figure{fig:retarder}.
Due to the spatial separation of the magnetic sublayers, such heterostructures can be constructed by sublayer engineering using the depositing/etching processes in SyAF \cite{lavrijsen_magnetic_2013,yang_domain_2015} or exfoliating processes in $2$D magnetic materials \cite{huang_layer-dependent_2017,gong_discovery_2017,xing_electric_2017,wang_raman_2016,lee_ising-type_2016}.

\emph{The $2/1/2$ structure: a circular spin wave polarizer.}
As shown in \Figure{fig:212}(b), a $2/1/2$ structure is obtained by removing a central portion of a bilayer SyAF. Since the central monolayer ferromagnet only allows one type of circular polarization to pass, the $2/1/2$ realizes the function of a circular spin wave polarizer. The allowed polarization in the central monolayer is determined by its sublayer-index: the $2/\check{1}/2$ structure is a left circular spin wave polarizer by allowing only the left circular polarization to transmit, while the structure $2/\hat{1}/2$ is a right circular spin wave polarizer.
This polarization-selective transmission behavior is verified by micromagnetic simulations based on the full LLG equation \Eq{eqn:LLG}, as well as Green function method \cite{lan_antiferromagnetic_2017} based on linearized spin wave equations \Eq{eqn:EOM_afm}.
\Figure{fig:212}(a) shows the transmission probabilities $T_\ssf{L/R}$ for the left and right circular polarzied spin wave through a $2/\check{1}/2$ structure: $T_\ssf{L}\approx 1$, $T_\ssf{R}\approx 0$.
Due to the spatial inversion symmetry, it is also found that $T_\ssf{L}\approx 0$, $T_\ssf{R}\approx 1$ for the $2/\hat{1}/2$ structure as indicated in the inset of \Figure{fig:212}(a).

We further verify the polarizing effect of the $2/1/2$ structure in \Figure{fig:212}(b) by micromagnetic simulations. As seen, when a linearly polarized spin wave, composed of equal left/right circular components, is injected from the left side of a $2/\check{1}/2$ structure,
the left circular component maintains its amplitude in the whole propagation processes, while the right circular component is blocked from the central $\check{1}$ region.
As a result, only the left circular polarization passes, indicating that $2/\check{1}/2$ is a left-circular spin wave polarizer.

\emph{The $2/3/2$ structure: a circular spin wave retarder.}
As shown in \Figure{fig:retarder}(b)(d), the $2/3/2$ structure is constructed by capping a portion of the bilayer SyAF by a third ferromagnet layer, whose magnetization is antiferromagnetically coupled to the upper layer of the SyAF.
The central trilayer structure hosts both left/right circular polarizations, but their dispersions are no longer degenerated and are shifted asymmetrically as shown in \Figure{fig:pols}(c).
As a result, the trilayer portion of the $2/3/2$ structure would induce relative phase delay between the left and right circular components. Consequently, the $2/3/2$ structure realizes the function of a circular spin wave retarder.
Such phase delay behavior of the $2/3/2$ structure is also verified by micromagnetic simulations and the Green function calculations. As shown in \Figure{fig:retarder}(a)(c), for the spin wave frequency above the gaps for both circular polarizations, the transmission probabilities through the $2/3/2$ structure for both polarizations approach unity: $T_\ssf{L}\approx T_\ssf{R}\approx 1$, but a relative phase between the left and right circular modes $\delta \phi_\ssf{LR}$ accumulates across the central trilayer region, and is proportional to the length of the central trilayer segment.
Depending on the magnetization direction (in $\mp \hbz$) of the capping layer, the relative phase $\delta \phi_\ssf{LR}$ can be either positive for $2/\check{3}/2$ (\Figure{fig:retarder}(a)) or negative for $2/\hat{3}/2$ (\Figure{fig:retarder}(c)).

The functionality of the circular spin wave retarder using a $2/3/2$ structure is further confirmed by micromagnetic simulations in \Figure{fig:retarder}(b)(d).
A linearly polarized spin wave of frequency $\omega/2\pi = 6.5 ~\mathrm{GHz}$ is injected from the left side of the $2/3/2$ structure. When passing the central $\check{3}$ (or $\hat{3}$) region, its left and right circular components develops relative phase delay. Due to the phase delay, the combined linear polarization direction steadily rotates counter-clockwisely/clockwisely. As the spin wave exits the trilayer region, the accumulated phase delay $\delta\phi_\ssf{LR} \approx 0.31 \pi$ (or $-0.29\pi$), and the linear polarization is rotated by $\delta\phi_\ssf{LR}/2 \approx 0.155\pi$ (or $0.145 \pi$).

\emph{Discussions.}
The present authors have previously reported a realization of spin wave polarizer and retarder based on antiferromagnetic domain walls \cite{lan_antiferromagnetic_2017}.
Apart from the completely different physical principles behind these two means of spin wave manipulation, the other most fundamental distinction is the basis for the spin wave polarization:
The sublayer-based polarization manipulation demonstrated here works in the basis of the left and right circular polarizations,
while the domain-wall based spin wave manipulation in Ref. \cite{lan_antiferromagnetic_2017} works in a basis of linear polarizations.
Being operating in different basis, the complementary schemes proposed here and that in Ref. \cite{lan_antiferromagnetic_2017} provide a more complete and flexible capabilities in harnessing spin wave polarization.

The sublayer engineering is not limited to the sublayer removing/capping as demonstrated here, but also includes other operations that introduce asymmetry between two sublayers, such as suppressing/enhancing the magnetization in one sublayer.
More generally, similar to the spin wave polarization-sublayer locking feature here, the sublayer may be also linked to other degrees of freedom such as valley or spin, thus these degrees of freedom locked to sublayer can be manipulated via similar sublayer engineering as presented here.

\emph{Conclusions.}
In conclusion, based on the spin wave polarization-sublayer locking feature, we propose to manipulate spin wave polarization via sublayer engineering  in SyAF, such as removing or capping a magnetic sublayer. A spin wave polarizer and retarder are realized in the basis of circular polarizations.
Controlling spin wave polarization by sublayer engineering in SyAF offers new possibilities to polarization-based magnetic information processing, and eventually the spin wave computing \cite{yu_spin_2018}.

\emph{Method.}
The micromagnetic simulations in this work are performed in COMSOL Multiphysics \cite{comsol}, where the LLG equation is transformed into weak form.
The SyAF wire is composed of two ferromagnetic YIG wires with following parameters \cite{yan_all-magnonic_2011,lan_spin-wave_2015,lan_antiferromagnetic_2017}:
the easy-axis anisotropy $K=8.57\ \mathrm{GHz}$, the exchange constant $A=7.25\times10^{-6}\ \mathrm{Hz}\cdot\mathrm{m^2}$, the inter-layer coupling $J=4.25\ \mathrm{GHz}$,
the gyromagnetic ratio $\gamma= 2.21 \times 10^{5} ~ \mathrm{Hz}/(\mathrm{A}/\mathrm{m})$, and the damping $\alpha=5\times 10^{-4}$.

\emph{Acknowledgements.}
This work was supported by the National Natural Science Foundation of China under Grant No. 11722430 and No. 11474065, National Key Research Program of China under Grant No. 2016YFA0300702, and National Basic Research Program of China under Grant No. 2014CB921600.

 \end{document}